\documentclass[twocolumn,prb,floatfix]{revtex4}
\usepackage{graphicx}    % Include figure files
\usepackage{dcolumn}     % Align table columns on decimal point
\usepackage{bm}          % bold math
\usepackage{amsmath}          % bold math
\begin{document}
%\preprint{Temperature} 

\title{Remarkable suppression of dc Josephson current 
on $d$-wave superconductor junction}
\author{Takashi Hirai }
\address{Nakatsugawa, gifu, 508-0101, Japan}
\date{\today}
\begin{abstract}
Josephson current in superconductor/insulator/superconductor junction
is studied theoretically. 
It is well known that 
when the zero-energy resonance state exists 
both side of superconducting interface, 
the behaver of the temperature dependence 
of the critical Josephson current is striking   
enhancement at low temperature.
On the other hand it is reported that 
if $d+is$-wave exists at the interface, 
Josephson current is suppressed at low temperature. 
In this paper, we discuss 
the existence of the imaginary part of the pair potential 
at the interface and 
remarkable suppresses of dc Josephson current 
on $d$-wave superconductor 110-junction.
\end{abstract}
\maketitle
\section{Introduction}
\indent 
In two decade, transport property of the 
unconventional superconducting junctions 
is studied both theoretically and experimentally. 
In these junctions, zero-energy resonance state (ZES) 
plays an important role. 
\cite{ZES2,ZES3}
It is well known that in the tunneling spectroscopy 
of the high-$T_C$ superconductor
the zero-bias conductance peak  
\cite{ZBCP4,ZBCP5,ZBCP6,ZBCP7,ZBCP8,ZBCP9,ZBCP10,ZBCP11} 
appears. 
\par 
On the other hand Josephson current 
in superconductor / insulator / superconductor junction 
is one of the characteristic phenomena. 
Anomalous behaver is obtained on high-$T_C$ 
superconducting junction, $i.e.$ 
the critical Josephson current 
enhances at low temperature 
when the lattice orientation is 
$\alpha =\pi /4$ (110-junction) as the Fig. 1. 
This is caused by the existence of the ZES 
formed at the interface.
In previous papers we have known a general formula 
as Furusaki-Tsukada formulation 
for dc Josephson current, 
which include both macroscopic phase and ZES. 
This theory is based on a microscopic 
calculation of the current represented 
in terms of the coefficients of Andreev reflection.
\cite{Andreev1,Andreev2,Andreev3} 
\par 
In this paper, we calculate and discuss dc Josephson current 
at 110-junction in the $d$-component superconductor / 
insulator / $d$-wave superconductor junction 
considering existence of $is$-wave state 
\cite{d+is} 
and imaginary part of the $d$-wave state.
In these, we calculate spatial dependence 
of the pair potential self-consistently. 
\par 
\begin{figure}[hob]
\begin{center}
%\rotatebox{270}
\includegraphics[width=8.0cm,clip]{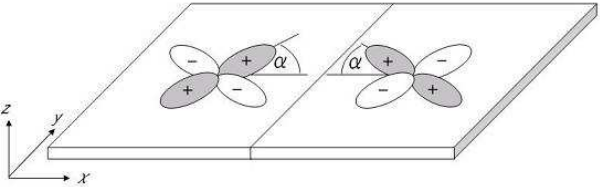}
\end{center}
\vskip -4mm
\caption{A schematic of 110-junction 
of the wave function of $d$-wave superconductors. 
The crystal orientation of 
right and left side of the superconductor for junction 
are chosen as $\alpha =\pi /4$, respectively. 
\label{fig:1}
} 
\end{figure}

\section{Formulation}
\indent 
In order to calculate Josephson current, 
we well know the Green's function method like this: 
\begin{eqnarray}
I=\frac{e\hbar}{2im}\left( \frac{\partial }{\partial x}
-\frac{\partial }{\partial x'} \right ) 
\mbox{Tr} G_{\omega_m}(x,x').
\end{eqnarray}
And now, we use the quasi-classical method in this paper.
First of all, Nanbu-Gol'kov Green's function is written as,
\begin{eqnarray}
\label{Green}
G(x,x')=G_{++}(x,x')e^{ik_F (x-x')}
       +G_{--}(x,x')e^{-ik_F (x-x')} \nonumber \\ 
       +G_{+-}(x,x')e^{ik_F (x+x')}
       +G_{-+}(x,x')e^{-ik_F (x+x')}. 
\end{eqnarray}
%
%here, 
%%
%\begin{eqnarray}
%G_{\alpha \beta}(x,x')
%=\frac{\phi_{\alpha n}(x)\phi_{\beta n}^{\dagger}(x')}
%{E-E_n}.
%\end{eqnarray}
%%
Taking the differential into the equation, 
we obtain the following equation, 
\begin{eqnarray}
I=\frac{e\hbar k_F}{m}\mbox{Tr}  
\left( 
G_{++}(x,x')e^{ik_F (x-x')}\right. \nonumber \\ 
\left. -G_{--}(x,x')e^{-ik_F (x-x')} 
\right) +O(1). 
\end{eqnarray}
The differential for $G_{\alpha\beta}(x,x')$ 
is order $1$ ( $\ll k_F$ ), 
so it's ignored. 
The quantity $\alpha$, $\beta$ mean $\pm$. 
The Green's function $G_{\pm \mp}(x,x')$ 
terms vanish in the differential. 
\par 
%
%The new defined Green's function obey the 
%Bogoliubov-de Gennes equations, 
%
%\begin{eqnarray}
%\left[ 
%E-\left( 
%\begin{array}{cc}
%-\alpha iv_F \partial _x & \Delta (k_F ,x)\\
%\Delta^{*} (k_F ,x) & \alpha iv_F \partial
%\end{array}
%\right) 
%\right] G_{\alpha \beta } (x,x')
%=\delta_{\alpha \beta } \delta (x-x'). 
%\end{eqnarray}
%
%Integrating the Green's function, we obtain 
%
%\begin{eqnarray}
%G_{\alpha\alpha}(x+0,x)-G_{\alpha\alpha}(x-0,x)
%=i\frac{\alpha}{|v_F|}\rho_{3}. 
%\end{eqnarray}
%
%To avoid the discontinuity of Green's function at $x=0$, 
We define the quasi-classical Green's function.
\cite{Green} 
%
%\begin{eqnarray}
%\hat{g}_{\alpha}\pm (\gamma_{3})
%=-2|v_F|\rho_{3}G_{\alpha\alpha}(x\pm 0,x).
%\end{eqnarray}
%
\begin{eqnarray}
\hat{g}_{\alpha}=f_{1\alpha}\hat{\tau}_1 
+f_{2\alpha}\hat{\tau}_2 
+g_{\alpha}\hat{\tau}_3 \mbox{  ,   } 
(\hat{g}_{\alpha})^2 =\hat{1}
\end{eqnarray}
Here $\hat{\tau}_j$($j=1,2,3$) are Pauri matrices and $\hat{1}$
is a unit matrix. 
The quantities $f_{1\alpha},f_{2\alpha},g_{\alpha}$ obey the 
following relations, 
%
%Changing the Green's function formula for 
%quasi-classical Green's function formula,
%we obtain
%
%\begin{eqnarray}
%I=\frac{e\hbar k_F}{m} \mbox{Tr}   
%\rho_3 \left[ 
%g_{+}(x)-g_{-}(x) 
%\right] . \nonumber \\ 
%\end{eqnarray}
%
%The quasi-classical Green's function for 
%$g_{\alpha\alpha}(x)$ is written down as
%%
%\begin{eqnarray}
%g_{++}(x,\theta )=i\left( 
%\begin{array}{cc}
%\frac{1+D_{+}(x)F_{+}(x)}{1-D_{+}(x)F_{+}((x)} & 
%\frac{2iF_{+}(x)}{1-D_{+}(x)F_{+}(x)} \\ 
%\frac{2iD_{+}(x)}{1-D_{+}(x)D_{+}(x)} &  
%-\frac{1+D_{+}(x)F_{+}(x)}{1-D_{+}(x)F_{+}(x)}
%\end{array}
%\right) ,\\ 
%
%g_{--}(x,\theta )=i\left( 
%\begin{array}{cc}
%\frac{1+D_{-}(x)F_{-}(x)}{-1+D_{-}(x)F_{-}(x)} & 
%\frac{2iF_{-}(x)}{-1+D_{-}(x)F_{-}(x)} \\ 
%\frac{2iD_{-}(x)}{-1+D_{-}(x)D_{-}(x)} &  
%-\frac{1+D_{-}(x)F_{-}(x)}{-1+D_{-}(x)F_{-}(x)}
%\end{array}
%\right) .
%\end{eqnarray}
%
%Here, $\theta$ is the angle between 
%quasi-particle going through the interface
%and $x$ direction, that is the vertical to the interface. 
%
\begin{eqnarray}
f_{1\alpha}=&\alpha 
\left[ i F_{\alpha}(x)+D_{\alpha}(x) \right] 
/\left[ 1-D_{\alpha}(x)F_{\alpha}(x) \right] ,\\
f_{2\alpha}=& 
-\left[ F_{\alpha}(x)-D_{\alpha}(x) \right] 
/\left[ 1-D_{\alpha}(x)F_{\alpha}(x) \right] ,\\
g_{\alpha}=& \alpha 
\left[1+ F_{\alpha}(x)D_{\alpha}(x) \right] 
/\left[ 1-D_{\alpha}(x)F_{\alpha}(x) \right] . 
\end{eqnarray}
In these quasi-classical Green's function, 
the quantity $D_{\alpha}(x)$ and $F_{\alpha}(x)$
obey the Ricatti equations \cite{Ricatti}
\begin{eqnarray}
\label{D}
\hbar |v_F |D_{\alpha}(x)
=\alpha \left[ 
2\omega_m D_{\alpha}(x)
+\Delta (x,\theta )D_{\alpha}^{2}(x) 
\right. \nonumber \\ 
 \left. 
-\Delta^{*}(x,\theta ) 
\right], \\ 
\hbar |v_F |F_{\alpha}(x)
=\alpha \left[ 
-2\omega_m F_{\alpha}(x)
+\Delta^{*} (x,\theta )F_{\alpha}^{2}(x) 
\right. \nonumber \\ \left. 
-\Delta(x,\theta ) 
\right].  
\end{eqnarray}
%
%here, the quantity $D_{\alpha}(x)$
%is defined as following:
%
%\begin{eqnarray}
%D_{\alpha}(x)
%=\frac{v_{\alpha}(x)}{u_{\alpha}(x)}e^{-i\eta}. 
%\end{eqnarray}
%\par 
%
The quantity $\theta$ is the angle between 
quasi-particle going through the interface
and $x$ direction, here $x$-axis is the vertical to the interface. 
The boundary conditions at the interface are given by
\begin{eqnarray}
\label{boundary11}
F_{+L}=\frac{D_{-R}-RD_{+R}-(1-R)D_{-R}}
{D_{-L}(RD_{-R}-D_{+R})+(1-R)D_{+R}D_{-R}}, \\ 
\label{boundary12}
F_{-L}=\frac{RD_{-R}-D_{+R}+(1-R)D_{+L}}
{D_{+L}(D_{-R}-RD_{+R})-(1-R)D_{+R}D_{-R}}, \\ 
\label{boundary13}
F_{+R}=\frac{RD_{+L}-D_{-L}+(1-R)D_{-R}}
{D_{-R}(D_{+L}-RD_{-L})-(1-R)D_{+L}D_{-L}}, \\ 
\label{boundary14}
F_{-R}=\frac{D_{+L}-RD_{-L}-(1-R)D_{+R}}
{D_{+R}(RD_{+L}-D_{-L})+(1-R)D_{+L}D_{-L}}, 
\end{eqnarray}
where we omit the index $(x=0)$. The quantity $R$ is 
$R=Z^2 /(4+Z^2 )$ with $Z=2mH/\hbar ^2 k_F$. 
Here the quantity $H$ is the height of the 
barrier potential.
Then we treat the insulator as the $\delta$-functional barrier potential. 
The boundary condition for $D_{\alpha}(x)$ at $x=\pm \infty$ is 
\begin{eqnarray} 
\label{boundary2}
D_{\alpha}(\pm \infty)
=\frac{\Delta^{*}(\pm \infty, \theta)}
{\omega_{m} +\alpha \Omega_{\alpha}}. 
\end{eqnarray}
\par 
In these relation, we can write down 
Josephson current as following, 
\begin{eqnarray}
\label{Josephson}
I(\theta )=\frac{2e\hbar k_F}{m} 
i\left( 
\left[ g_{+}(x,\theta ) \right]
-\left[ g_{-}(x,\theta ) \right]
\right). 
\end{eqnarray}
Josephson current in this formula 
is obtained by $x\rightarrow 0$
\par 
The spatial dependent pair potential 
is calculated as following
%
%\begin{eqnarray}
%\label{pair}
%\Delta(x,\theta )&=&\frac{2\pi k_B T}
%{\log T/T_C +\sum_{0\leq m}\frac{1}{m+1/2}}
%\sum_{0\leq m}\frac{1}{2\pi}
%\int_{-\pi /2}^{\pi /2}
%\frac{d\theta '}{2\pi} \nonumber \\ 
%&\times &\left[ 
%V(\theta ,\theta '_{+} ) 
%\left\{
%\left[ g_{++}(\theta '_{+}) \right]_{12}
%-\left[ g_{++}^{\dagger}(\theta '_{+}) \right]_{12}
%\right\}
%\right. \nonumber \\ 
%&+&\left. 
%V(\theta ,\theta '_{-} )  
%\left\{ 
%\left[ g_{--}(\theta '_{-}) \right]_{12}
%-\left[ g_{--}^{\dagger}(\theta '_{-}) \right]_{12}
%\right\} 
%\right].\nonumber \\ 
%\end{eqnarray}

\begin{eqnarray}
\label{pair}
\Delta(x,\theta )&=&\frac{2T}
{\ln T/T_C +\sum_{0\leq m}\frac{1}{m+1/2}}
\nonumber \\ 
&\times &\sum_{0\leq m}
\int_{-\pi /2}^{\pi /2} 
d\theta ' 
V(\theta ,\theta ' ) f_{2+} 
\end{eqnarray}
where $V(\theta ,\theta ' )=2\sin 2\theta \sin 2\theta '$
for 110-junction 
and $V(\theta ,\theta ' )=2\cos 2\theta \cos 2\theta '$
for 100- junction, respectively for $d$-wave component, 
and $V(\theta ,\theta ' )=1$ for $s$-wave component 
for both 110- and 100-junction case.
In this equation, we can calculate spatial dependent of 
the pair potential self-consistently (SCF).  
\par 
Josephson current $I$ in these formula 
is obtained numerically solving Eq. 
\ref{D}, \ref{Josephson}, \ref{pair}
under the boundary conditions Eq. 
\ref{boundary11},\ref{boundary12},\ref{boundary13},\ref{boundary14},\ref{boundary2}. 
\par 
Calculated result of Josephson current 
is normalized by normal conductance $\sigma_N$, 
\begin{eqnarray} 
I(\eta )=\int_{-\pi /2}^{\pi /2}I(\theta )\cos\theta d\theta
/\sigma_N, 
\end{eqnarray} 
\begin{eqnarray}
\sigma_N =\int_{-\pi /2}^{\pi /2} \frac{4\cos \theta^2}{4\cos \theta^2+Z^2} 
\cos \theta d\theta .
\end{eqnarray} 
Here, we define $\eta =\eta_L -\eta_R$, 
where $\eta_L$, $\eta_R$ is the 
macroscopic phase of left and right side of the superconductors. 
In every thing, we chose the temperature $T=0.05T_C$, 
where $T_C$ is the transition temperature of superconductor. 
And the cutoff frequency $\omega_D$ is set to be $\omega_D /2\pi T_C =1$ 
for summation of Matsubara frequency $m$. 
\section{Results}
In this section, we show the calculated results 
of the superconducting macroscopic phase $\eta$ 
dependence of the pair potential and Josephson current. 
In all case we chose $T_{C_s}=0.2T_{C_d}$, 
where $T_{C_s}$ and $T_{C_d}$ ($=T_C$) are the transition temperature
of $s$-wave component and $d$-wave component, respectively. 
\par
\begin{figure}[hob]
\begin{center}
%\rotatebox{270}
\includegraphics[width=5cm,clip]{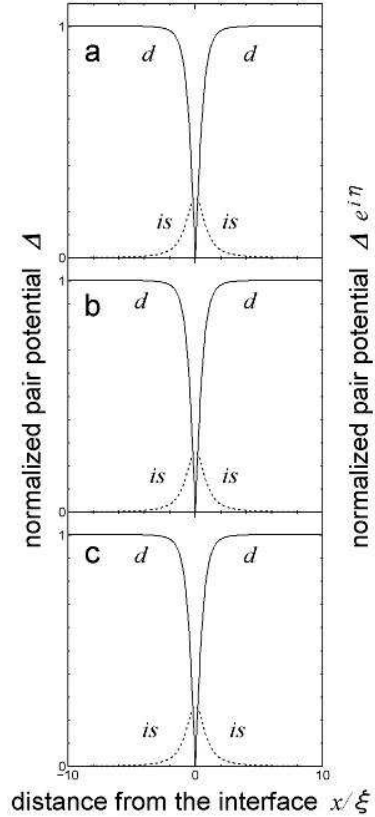}
\end{center}
\vskip -4mm
\caption{The $x$-dependence of the 
pair potential of the right and left side of the 
superconductors at $\eta =0$. 
The a, b, c mean $Z=0$, $5$, $10$, respectively. 
$\xi$ is the coherent length of the superconductor.
\label{fig:2}
} 
\end{figure}
\begin{figure}[hob]
\begin{center}
%\rotatebox{270}
\includegraphics[width=5cm,clip]{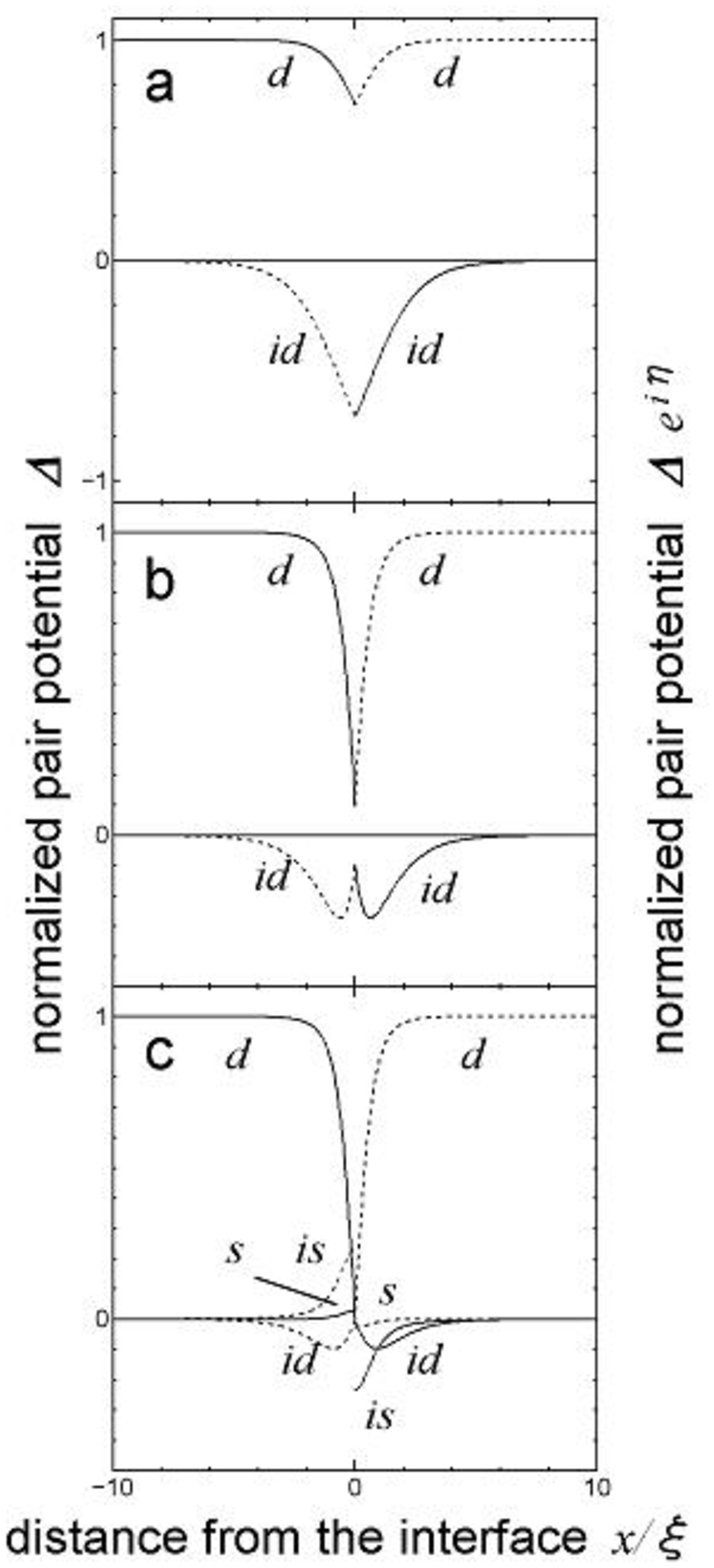}
\end{center}
\vskip -4mm
\caption{The $x$-dependence of the 
pair potential of the right and left side of the 
superconductors at $\eta =\pi /2$. 
The a, b, c mean $Z=0$, $5$, $10$, respectively.  
$\xi$ is the coherent length of the superconductor.
\label{fig:3}
} 
\end{figure}

\begin{figure}[hob]
\begin{center}
%\rotatebox{270}
\includegraphics[width=4.8cm,clip]{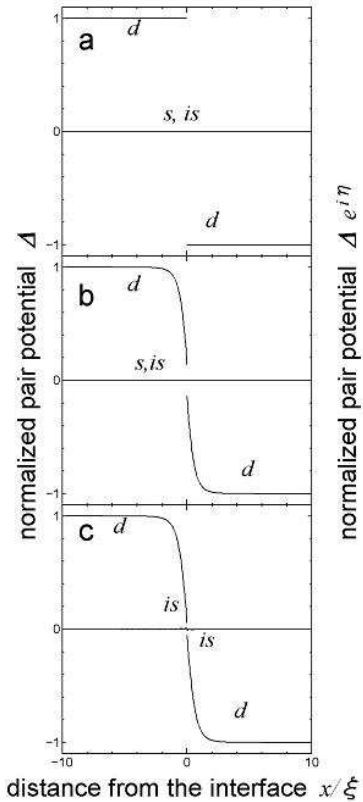}
\end{center}
\vskip -4mm
\caption{The $x$-dependence of the 
pair potential of the right and left side of the 
superconductors at $\eta =\pi$. 
The a, b, c mean $Z=0$, $5$, $10$, respectively.  
$\xi$ is the coherent length of the superconductor.
\label{fig:4}
} 
\end{figure}

First of all, we show the $\eta$-dependence of the 
pair potential. 
The system is 110-junction.
In the Fig. 2, 3, 4, 
the solid line is real number, 
and doted line is imaginary number, respectively. 
The $x$ axis is normalized by coherent length $\xi $. 
For $\eta =0$, $Z$-dependences 
don't appear in the pair potential as Fig. 2. 
The reducing of the pair potential for $Z=0$ near the interface 
receives spatial changing. 
In these, $is$ state exists near the interface. 
This state doesn't appear in the 100-junction's case. 
%where $\alpha$ is the angle between 
%the crystal lattice and the interface. 
When the quasi-particle goes through the interface,
it feels the opposite sign of the pair potential.
So the reducing occurs in spite of $Z=0$. 
\par 
Second, let us show the $\eta =\pi /2$ case. 
For $Z=0$, $s$ and $is$ state not exist. 
Since the pair potential contains 
the superconducting macroscopic phase 
at the right side of the superconductor, 
in the Fig. 3, right side of the pair potential 
of $d$-wave is imaginary number. 
Similarly  $\eta =0$ case, 
since the quasi-particle through the interface 
feels the different sign of the pair potential, 
real number of $d$-wave 
(since it contains the macroscopic phase of the superconductor, 
real number of $d$-wave appears as the imaginary number. )
at the right side of the pair potential  
is connected to the imaginary $d$-wave 
at the left side of the pair potential. 
\par 
For $Z=5$, pair potential behaves as Fig. 3 $b$.
The existence of the barrier potential 
affects the suppression of the right and left side 
of the pair potential for both real and imaginary numbers 
near the interface. 
At the region of the coherence length near the interface,  
imaginary parts of the $d$-wave is enhanced 
for the right and left side of the superconductors. 
\par 
When $Z=10$, $s$ and $is$ state appear 
for both side of the superconductors 
near the interface. 
The existence of the $is$-wave 
is same reason as the ordinary discussion for 
$is$-wave state at the edge of the $d$-wave superconductor
on $\alpha =\pi /4$ (110-junction). 
\par 
Next, we show the $\eta =\pi$ case. 
The phase factor is $\exp (i\eta)=-1$,
so 110-junction is same as 
in the 100-junction ($\alpha =0$). 
Therefore real part of the $d$-wave 
is not spacial dependence in the $Z=0$ case, 
$i.e.$ pair potential is constance for all region. 
$s$ and $is$-wave don't appear 
and $id$-wave doesn't appear too. 
\par 
When $Z=5$, since existence of the barrier potential 
makes the reflection of the quasi-particle 
at the interface, 
$d$-wave factors are reduced. 
$Z=10$ case is same as $Z=5$. 
The different point at the $Z=10$ is 
the existence of the $is$ state at the interface. 
\par 
Finally, we show the normalized dc Josephson current 
for 110-junction and 100-junction.
For 110-junction, 
Josephson current is suppressed at the 
$\eta =\pi /2 \sim \pi$ region for $Z=10$, 
and it suppressed at all area for $Z=5$. 
For $Z=15$, Josephson current behaves $\sin \eta$. 
On the other hand, for 100-junction, 
Josephson current is not suppressed. 
And it is consistent with the non-SCF calculation. 
\par 
Comparing the 110-junction to 100-junction, 
the hight of the Josephson current for 100-junction 
is higher than that for 110-junction. 
It is not consistent with non-SCF calculations. 
This is our new dissolve. 
\section{Summary} 
In this section, we summarize the obtained results. 
Now we have seen the imaginary part of the pair potential exists 
at the $\eta \neq 0$, $\pi $ for 110-junction. 
That occurs by the existence of the macroscopic phase 
of the superconductors. 
This results are different from the situation of the surface of superconductor 
or junction between normal metal and superconductor. 
Since the both $is$- and $id$-wave state exist near the interface, 
Josephson current is reduced on 110-junction. 
This results is not only by the $is$-wave state 
but also by the existence 
of the imaginary part of the pair potential of $d$-wave. 
This reducing is same as in the 
$s$-wave superconductor / $p$-wave-superconductor / $s$-wave superconductor junction.
In this junction similar reducing occurs by the existence 
of another symmetry pair potential at the junction. 
\cite{Yamashiro} 
In this paper's case $id$-wave component plays 
the different symmetry for the $d$-wave component.
On the other hand, Josephson current is not reduced 
on 100-interface. 
These results are unusual. 
These appear only in the SCF calculation. 
In the non-SCF calculation, 
these don't appear. 
These result for 110-junction is consistent with Ref. 10, 
where it's a high barrier limit case. 
\par 
And adding one more thing, 
Josephson current disappears on  $Z=0$ both for 110-junction 
and 100-junction. 
%This result means the $Z=0$  
%is not a physical situation. 
%The situation that barrier potential is nothing 
%and macroscopic phase of the 
%right and left side of the superconductors are not same
%is unnatural. 
%
In the physical point of view, 
it is expected that Josephson current only exists 
when insulating barrier or something 
(normal metal or different type of superconductor) 
exist at the interface. 
Therefore these results are valid physically. 
%
%This results just appears in the SCF calculation, too. 
%These are suitable for experimental results. 
\par 
In this paper, we discuss dc Josephson current 
for the 110-junction. 
Pair potential has the imaginary part for $\eta \neq 0$, 
and Josephson current is suppressed. 
This result appears only in the SCF calculation of the pair potential. 
\begin{figure}[hob]
\begin{center}
%\rotatebox{270}
\includegraphics[width=4.5cm,clip]{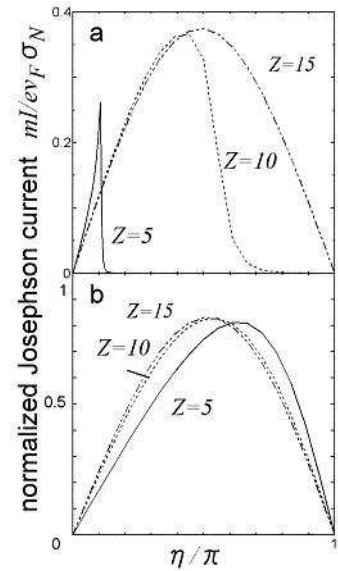}
\end{center}
\vskip -4mm
\caption{Josephson current at the 110-junction (a) 
and 100-junction (b) for $Z=5$, $10$ and $Z=15$. 
\label{fig:5}
} 
\end{figure}

\acknowledgements 
I greatly acknowledge useful comment with Y. Tanaka and N. Hayashi. 
I would like to thank S. Kaya  for giving me a calculating tool.

\end{document}